\documentclass[conference]{IEEEtran}

\usepackage{cite}
\usepackage{amsmath,amssymb,amsfonts}
\usepackage{algorithmic}
\usepackage{graphicx}
\usepackage{textcomp}
\usepackage{booktabs}
\usepackage{enumerate}
\usepackage{caption}
\usepackage{enumitem}
\usepackage{multirow}
\usepackage{subcaption}

\usepackage[colorlinks=true,citecolor=blue]{hyperref}

\usepackage[dvipsnames]{xcolor}
\usepackage{colortbl}
\definecolor{lightblue}{HTML}{E0ECFF}

\usepackage{amsmath,amsfonts,bm}

\def\eqref#1{equation~\ref{#1}}

\def\1{\bm{1}}

\def\vx{{\bm{x}}}
\def\vy{{\bm{y}}}

\DeclareMathAlphabet{\mathsfit}{\encodingdefault}{\sfdefault}{m}{sl}
\SetMathAlphabet{\mathsfit}{bold}{\encodingdefault}{\sfdefault}{bx}{n}

\def\BibTeX{{\rm B\kern-.05em{\sc i\kern-.025em b}\kern-.08em
    T\kern-.1667em\lower.7ex\hbox{E}\kern-.125emX}}

\newcommand\blfootnote[1]{%
  \begingroup
  \renewcommand\thefootnote{}\footnote{#1}%
  \addtocounter{footnote}{-1}%
  \endgroup
}

\begin{document}

\title{\title{Speech Recognition with LLMs Adapted to Disordered Speech Using Reinforcement Learning}
}

\author{
\IEEEauthorblockN{
Chirag Nagpal$^{*}$ \ \
Subhashini Venugopalan$^{*}$ \ \
Jimmy Tobin \ \
Marilyn Ladewig \\
Katherine Heller \ \ 
Katrin Tomanek$^{\dagger}$ 
}
\IEEEauthorblockA{\textit{Google Research} \\
Mountain View, USA 
}
}

\maketitle

\begin{abstract}

We introduce a large language model (LLM) capable of processing speech inputs and show that tuning it further with reinforcement learning on human preference (RLHF) %
enables it to adapt better to disordered speech %
than traditional fine-tuning. %
Our method replaces low-frequency text tokens in an LLM's vocabulary with audio tokens and enables the model to recognize speech by fine-tuning it on speech with transcripts. 
We then use RL with rewards based on syntactic and semantic accuracy measures  %
generalizing the LLM further
to recognize disordered speech. 
While the resulting LLM does not outperform existing systems for speech recognition we find that tuning with reinforcement learning using custom rewards leads to substantially better performance than supervised fine-tuning of the language model, specifically when adapting to speech in a different setting. %
This presents a compelling alternative tuning strategy for speech recognition using large language models.
\blfootnote{$^\star$equal contribution, $^\dagger$lead}

\end{abstract}


\section{Introduction}
Recent research has advanced multimodal potential of large language models (LLM) improving their visual processing capabilities~\cite{dubey2024llama,beyer2024paligemma,team2023gemini,liu2024visual} in addition to their knowledge and language understanding. Increasing the capabilities of LLMs to process audio while retaining their linguistic capabilities~\cite{zhang2023speechgpt,rubenstein2023audiopalm,fathullah2024prompting} will enable further interesting applications particularly with voice controlled automation that can encompass additional contextual cues. %
The ability to recognize speech in conjunction with the additional visual and textual context that LLMs can process 
could be particularly beneficial to people with accessibility needs such as individuals with speech impairments. 
A step in this direction is to investigate strategies to enable automatic speech recognition (ASR) using LLMs.  Here, our goal is two-fold, we want to enable training an existing LLM to recognize speech inputs, further, we investigate strategies to adapt the model to recognize disordered speech.

\begin{figure*}[!t]
    \centering
    \includegraphics[width=0.85\textwidth]{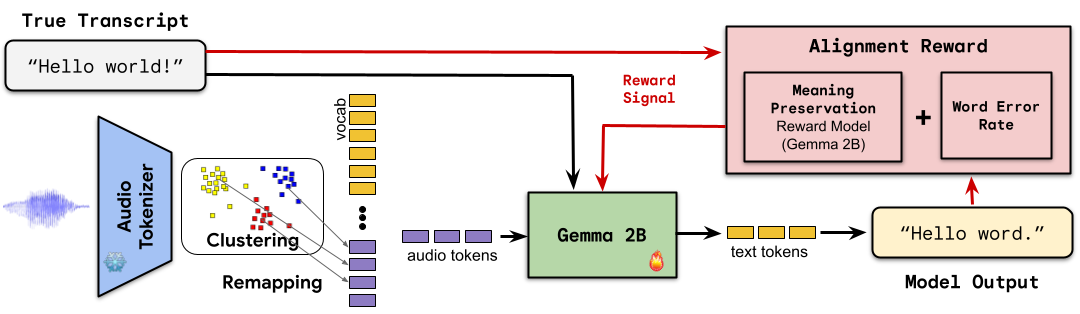}
    \caption{\textbf{Using an LLM for ASR}: Our approach involves first clustering the audio embeddings to the LLMs vocabulary space. Followed by supervised finetuning on a mixture of disoredred speech. In the final step we use reinforcement learning to improve the model's output in terms of Word Error Rate and human assessed meaning preservation quality.}
    \vspace{-10pt}
    \label{fig:asr_llm}
\end{figure*}

\textbf{LLM-based ASR.}
While recent work has shown how the architecture of larger LLMs can be modified~\cite{rubenstein2023audiopalm,fathullah2024prompting} and trained to recognize speech comparable to state of the art ASR models, it is still unclear if 
this is achievable with 
smaller LLMs and whether such models can be adapted further e.g., in a low resource setting. Many early works look at revising the outputs of ASR systems using LLMs~\cite{chen2023large,zheng2021adapting,xu2022rescorebert} but only a few have started looking at tuning an LLM directly to recognize speech. Most of these~\cite{fathullah2024prompting,wu2023decoder,ma2024embarrassingly} work by utilizing a speech encoder to extract embeddings, and often compress or learn 
a linear projection to the language model's embedding space, and then tune the model to recognize speech. 
Our work is more similar to~\cite{zhang2023speechgpt} where they discretize the audio embeddings and use the discrete units as input rather than directly training from the embeddings.
However, %
while~\cite{zhang2023speechgpt} expand the 
vocabulary of the LLM prior to tuning it, we simply repurpose low frequency text tokens in the vocabulary for audio and tune the LLM to enable the model to recognize speech. %

\textbf{Semantic distance measures for ASR.} To adapt the LLM based ASR system to specific domains, we propose tuning of the model using RL with a reward towards preserving semantics of the speech. Established ASR systems typically rely on quantitative scalar metrics such as Word Error Rate (WER)   for assessment of syntactic accuracy. However, in some cases, even when word error rates are relatively low, the semantic changes introduced by the errors can critically affect the intended meaning, and fall short on determining comprehensibility. e.g. ``Not so good, okay.'' vs.~``That's so good, okay''.  This issue is exacerbated in the case of disordered speech where ASR systems have higher than normal error rates. To mitigate such issues, \cite{tobin2022assessing,kim2021semantic} propose methods that learn semantic distance measures to compare ASR systems. \cite{shor2023clinical,lattescore_icassp2024} take it a step further to build systems that use human preference, i.e. clinician or speech language pathologist labels to determine whether transcripts preserve meaning.
While \cite{shor2023clinical} %
proposes a modification to re-weight terms when computing the existing BERTScore~\cite{zhang2019bertscore} metric, %
~\cite{lattescore_icassp2024} propose an approach training an LLM as a judge to evaluate if a transcript preserves the intended meaning. In this work, we evaluate the use of both WER and the LLM Meaning Preservation (MP) judge as rewards towards adapting an LLM-based ASR system on disordered speech, and show it to outperform fine-tuning.

\section{LLMs for ASR}

For our first stage %
we tune an LLM to function as an ASR model. Specifically, we enable a decoder-only transformer based LLM to take text and audio tokens as input and generate text corresponding to the transcript. This is made possible from the observation that the inputs to an LLM are essentially sequences of arbitrary integers obtained when text is tokenized and mapped to a vocabulary. In a similar manner, we first tokenize the audio using an audio tokenizer and discretize it to a very small finite vocabulary. We further map the audio tokens to replace %
text tokens in the language model's vocabulary. This setting while introducing no difference to the usual decoder-only setup for pure text is now capable of processing multimodal data as some of the tokens in the vocabulary now represent audio (Fig.~\ref{fig:asr_llm}). This model can then be tuned as a regular LLM without any changes to the architecture or training infrastructure. We describe these aspects below. %

\subsection{Using Audio Tokens in a standard LLM}
The first step towards modifying the LLM to understand speech is to tokenize the speech. Our goal is to retain the architecture of the LLM as-is while allowing it to also handle audio. Since the LLM first applies a tokenizer such as a sentence-piece~\cite{kudo2018sentencepiece} or word-piece~\cite{devlin2018bert} tokenizer to break text into a fixed vocabulary and takes discrete indices in the vocabulary as input, our key idea is to simply replace certain tokens in the vocabulary with audio tokens.

\textbf{Tokenizing and discretizing audio.} 
We use USM speech encoder~\cite{zhang2023google} trained similarly to w2v-BERT~\cite{zhang2023google} and produces tokens at the rate of 25Hz. We extract embeddings from the middle layers (16$^{th}$ layer) and cluster the embeddings into 1024 clusters.
Pilot experiments of varying layers, cluster sizes showed less difference.
Given a new speech sample, we first encode the audio into embeddings and identify the closest cluster center corresponding to each audio embedding, thus mapping the audio tokens to a sequence of cluster ids.  %

\textbf{Re-mapping text tokens as audio tokens.} %
Next, we replace some of the text tokens in the LLM vocabulary with audio tokens. We do that by mapping the 1024 audio cluster ids to the last 1024 vocabulary indices corresponding to the least frequent text-tokens.  This design choice is motivated from the fact that the last few tokens typically represent low frequency multilingual or unicode characters that can be re-purposed as our audio tokens. Once remapped, we perform standard supervised fine-tuning~\cite{team2024gemma} on ASR data where the inputs are the discretized audio tokens and output is the corresponding text transcript. %

\subsection{Supervised Finetuning an Audio LLM}
Since our goal is to not only train the model to recognize speech but also adapt it to more diverse disordered speech, we compare fine-tuning of the LLM to recognize speech on different mixtures of standard speech datasets and disordered speech data. With the tokenized audio data we use standard fine tuning recipes to tune the model. We first describe the specific model we use and then share parameters of the finetuning recipe experimented.

\section{Model and dataset}

\textbf{Using Gemma 2B for ASR} For the base LLM we use the open-source Gemma 2B~\cite{team2024gemma} model which is a decoder only transformer \cite{vaswani2017attention} with Multi Query Attention \cite{shazeer2019fast}. %
The Gemma~\cite{team2024gemma} model family vocabulary consists of 256k tokens, and %
replacing the last 1024 tokens in the vocabulary should not affect the underlying LLM capabilities in understanding or generation.

\noindent\textbf{Datasets.}
We use two datasets of English speech in our study. 
\begin{itemize}[leftmargin=*]
\item[-]{\textbf{LibriSpeech}}~\cite{librispeech} is a large-scale dataset consisting of training data of a 1000 hours standard speech from English audio books read by single speakers in a clean environment. We use the dev-clean split of the data for validation.

\item[-]{\textbf{Euphonia}}~\cite{macdonald2021disordered} is a dataset of disordered speech of over 1M utterances ($\sim$1k hrs) on everyday topics from speakers with diverse speech impairments. We use the SI splits described in~\cite{green2021automatic} consisting of over 900k utterances from 1246 speakers for training, a test set of 5699 utterances from 200 speakers, and a small validation set of 343 utterances from 24 speakers. The splits do not overlap in speakers or in phrases.

\end{itemize}

\noindent\textbf{Tuning details.}
Prior to training, we tokenize the audio datasets. We learn the clusters for the audio tokens based only on the LibriSpeech dataset. With the tokenized audio, we perform supervised finetuning of the Gemma model with a learning rate of $5\times10^{-6}$, and a cosine decay schedule, using the Adam optimizer. We experimented with different input drop outs, we found $5\times10^{-2}$ to be the best value.  Since our goal is to improve performance on disordered speech, we also experiment tuning of the LLM with different mixtures of the datasets. We experiment training the model on Librispeech-only, a 50:50 mix, and a 30:70 mix of Euphonia and Librispeech.   %
We evaluated the model on a held out validation set of both the Librispeech and Euphonia datasets and used the checkpoint with the lowest WER as our SFT model. %

\section{Domain adaptation with RLHF}
\label{sec:rlhf}

Our eventual goal is to adapt the LLM based ASR model to recognize disordered speech. This can be done by tuning the model further on the disordered speech data. %
In their seminal work \cite{ouyang2022training} demonstarted that RL can be used to finetune LLM outputs on a relatively small number of examples to maximize metrics learnt from human feedback. Motivated by this we %
explore tuning the model with a reward
from two sources of feedback, 
(1) \textbf{WER} between the transcript and ground truth, this provides a syntactic measure of accuracy of the prediction; and (2) \textbf{Meaning Preservation (MP)}~\cite{lattescore_icassp2024}, an LLM-based semantic measure of correctness.

{\textbf{Meaning Preservation reward model}}. In \cite{lattescore_icassp2024}, an LLM is trained specifically to mimic human judgements on whether ASR system predicted transcript preserves meaning compared to the groundtruth. While they use a 62B parameter LLM for this task, we train a Gemma-2B reward model on the same binary classification task of meaning preservation introduced in ~\cite{tobin2024automatic} on 2840 pairs of predicted and gold transcripts using a cross-entropy loss on the 0-1 labels. On a test set of about 1k examples our model achieved a 0.87 AUC, compared to the 0.89 AUC obtained by \cite{lattescore_icassp2024}.

\textbf{WER signal}. Several prior works on RLHF have found that RL training is especially sensitive to exploiting spurious correlations in the reward model leading to a phenomenon known as \textit{`reward hacking'} \cite{eisenstein2023helping, gao2023scaling, pang2022reward, wang2024transforming}. To mitigate this we also include WER metric as an additional signal.
This ensures that the model is rewarded to generate responses with low WER while maximizing meaning preservation score. 

\noindent The final reward signal $R$ used for alignment is given by Eqn.~\ref{eqn:reward}
\begin{equation}
R(\vx, \vy; \vy^*) := \gamma \cdot \texttt{MP}(\vy, \vy^*) + \ln \bigg( 1 - \texttt{WER}(\vy, \vy^*) \bigg)
\label{eqn:reward}
\end{equation}
here $\vx$ is the original prompt, $\vy$ is the predicted transcript, $\vy^*$ the ground truth transcript, %
and $\gamma$ is a hyperparameter to trade-off the relative weight of the WER and meaning preservation score. Note that in this final reward function, the WER is logarithmically transformed to have the same scale as MP.

\textbf{Reinforcement Learning}. We use Proximal Policy Optimization (PPO) with a clipped gradient objective and  KL regularization as proposed in \cite{schulman2017proximal} for optimizing the combined reward in Equation \ref{eqn:reward}. We experiment with different values for $\gamma$ and select checkpoints based on WER and MP scores on the validation set. We report results on val. and test splits. %

\section{Results and Discussion}

\subsection{Making an LLM ASR model}

\begin{figure}[!tb]
    \centering
    \textbf{Validation Loss on Disordered Speech (Euphonia)}\\
    \includegraphics[width=0.375\textwidth]{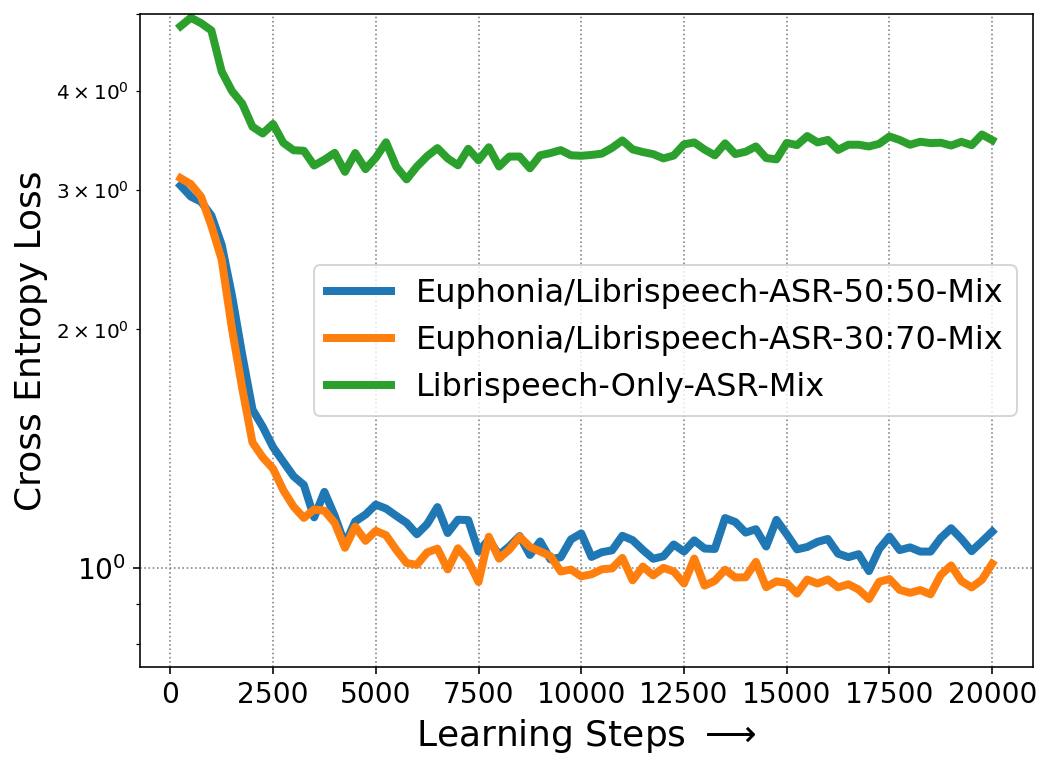}%
    \caption{Cross entropy Loss vs. Learning Steps of the Gemma 2B base LLM for different training mixtures of Librispeech and Euphonia data on a held-out validation set. Augmenting the disordered speech training data with generic ASR data from Librispeech improves generalization.}
    \label{fig:mixing}
\end{figure}

For training the base LLM-ASR model, Fig. \ref{fig:mixing} compares curves of the three mixtures of Euphonia:Librispeech data. We find that augmenting training with the disordered speech data, specifically the 30:70 mixture (Table~\ref{tab:gemma_asr}) shows significantly improved performance on the disordered speech data while retaining performance on the standard Librispeech dev-clean set. Training only on the Librispeech data or only the Euphonia datasets leads to extremely poor generalization highlighting the substantial differences in the characteristics of the datasets. %

\begin{table}[!ht]
    \centering
\small
\resizebox{\columnwidth}{!}{
\begin{tabular}{l|c|c|c|c|c|c}
        \toprule \midrule
        \multirow{2}{*}{\bf Dataset mixture} & \multicolumn{2}{c|}{\textbf{Euphonia  Test}} & \multicolumn{2}{c|}{\textbf{Euphonia  Dev}} &  \multicolumn{2}{c}{\textbf{Librispeech Dev}}\\ 
          & \textbf{WER $\downarrow$} & \textbf{MP $\uparrow$}  & \textbf{WER $\downarrow$} & \textbf{MP $\uparrow$} & \textbf{WER $\downarrow$} & \textbf{MP $\uparrow$}  \\
        \midrule
        Librispeech Only  & 70.9 \ & 39.0 \ & 66.5 \ & 31.8 \ & \textbf{17.1} & \textbf{86.6} \\
        30:70 mixture & \cellcolor{lightblue}\textbf{50.4}* & \cellcolor{lightblue}\textbf{48.2}* & \cellcolor{lightblue}\textbf{47.3}* & \cellcolor{lightblue}\textbf{48.1}* & 17.2 & 85.6 \\
        \bottomrule
    \end{tabular}
}
    \caption{Training the LLM on ASR data with a 30:70 mix of Euphonia:Librispeech leads to significant (\colorbox{lightblue}{*}) improvements on Euphonia and little loss on Librispeech. $\uparrow$ and $\downarrow$ indicate higher or lower is better respectively. \textbf{bold} shows best score.}
    \label{tab:gemma_asr}
\end{table}

\begin{table*}[!t]
    \centering
    \caption{Examples selected based on human evaluation of transcripts on meaning preservation and error type of the RLHF models show that trading-off WER slightly for a significant gain in MP score ($\gamma=1.00$) leads to better predictions overall. }
\resizebox{\textwidth}{!}{

    \begin{tabular}{c|c|l l|l l} \toprule \midrule
        \textbf{Ground Truth} & \textbf{Severity} & \textbf{RLHF} ($\gamma=0.0)$  & WER & \textbf{RLHF}  ($\gamma=1.0)$ & WER\\ \midrule
        \textsf{"not so good today"} & \texttt{MILD} & \textsf{"not so good to the."} & (0.5) & \textsf{"not so good to day."} & (0.5) \\
\textsf{"every one of my family listens to music"} &  \texttt{MODERATE} & \textsf{``every once in my frame and listen to music''} & (0.62) & \textsf{``everybody in my family listens to music''} & (0.38) \\
\textsf{``dancing is so much fun''} & \texttt{MODERATE} &\textsf{``that's so much fun.''} & (0.40) & \textsf{``dancing so much fun.''} & (0.20) \\
\textsf{``are you comfortable?''} & \texttt{MODERATE} &\textsf{``are you going to school?''} & (1.0) & \textsf{``are you comfortable with it?''} & (0.67) \\
\textsf{``happy birthday dear friend.''} & \texttt{SEVERE} &\textsf{``absolutely your friend.''} & (0.75) & \textsf{``happy birthday to your friend.''} & (0.50) \\
\textsf{``as soon as possible''} & \texttt{SEVERE} & \textsf{``it soon adds pounds him volume''} & (1.0) & \textsf{``a soon as possible.''} & (0.25) \\ \bottomrule
    \end{tabular}
}
    \vspace{-8pt}
    \label{tab:examples_rlhf_wer_mp}
\end{table*}

\subsection{Adaptation with RLHF}
\label{sec:res_rlhf}
In Table~\ref{tab:gemma_rlhf} we compare both continued fine-tuning and RL based on reward signals for tuning and adapting the LLM-ASR model on disordered speech. %
We find that both WER and meaning preservation provide strong signals for adaptation. Even with significant hyper parameter tuning we find that fine-tuning directly on the disordered speech data doesn't help at all, whereas RLHF (all values of $\gamma$) show significant improvements ($p<1\times10^{-8}$) relative to the SFT 30:70 mixture model which is the starting point for RLHF. Significance is computed using two-sided t-tests.

\begin{table}[!htbp]
    \centering
        \caption{Adapting the LLM based ASR model to disordered speech on Euphonia with fine-tuning and RLHF. $\uparrow$ and $\downarrow$ indicate higher or lower is better respectively. \textbf{bold} shows best score, and \colorbox{lightblue}{*} is significant w.r.t \texttt{\textbf{WER}}  $(\gamma=0.00)$ strategy.}

\resizebox{\columnwidth}{!}{
\begin{tabular}{r|c|c|c|c|c|c}
        \toprule \midrule
        \multirow{2}{*}{\textbf{Tuning strategy}} & \multicolumn{2}{c|}{\textbf{Euphonia  Test}} & \multicolumn{2}{c|}{\textbf{Euphonia  Dev}} &  \multicolumn{2}{c}{\textbf{Librispeech Dev}}\\ 
          & \textbf{WER $\downarrow$} & \textbf{MP $\uparrow$}  & \textbf{WER $\downarrow$} & \textbf{MP $\uparrow$} & \textbf{WER $\downarrow$} & \textbf{MP $\uparrow$}  \\
         \midrule
        Base SFT model & 50.4 & 48.2 \  & 47.3 & 48.1 \  & 17.2 & 85.6 \ \\

        Continued SFT & 57.1 & 42.8 \  & 59.2 & 40.5 \ & 22.9 & 73.2 \ \\
        \midrule
        \multicolumn{7}{c}{RLHF \texttt{\textbf{WER}} + \texttt{\textbf{MP}}} \\ %
        \midrule
        \texttt{\textbf{WER}}  $(\gamma=0.00)$ & \textbf{41.0} & 50.4 \ & \textbf{40.1} & 47.5 \ & \textbf{20.2} & 75.7 \ \\
          + \texttt{\textbf{MP}} $(\gamma=0.25)$ & 41.7 & 51.3 \ & 41.7 & 48.7 \ & 22.4 & 74.7 \ \\
          + \texttt{\textbf{MP}} $(\gamma=0.50)$ & 41.2 & 52.9 \ & 41.1 & 49.0 \ & 23.9 & 72.2 \ \\
          + \texttt{\textbf{MP}} $(\gamma=1.00)$ & 42.6 & \cellcolor{lightblue}\textbf{55.7}* & 42.9 & \cellcolor{lightblue}\textbf{52.5}* & 22.0 & \cellcolor{lightblue}\textbf{76.2}* \\
        \bottomrule
    \end{tabular}
}
    \label{tab:gemma_rlhf}
\end{table}

In particular we find that setting equal weight to WER and the meaning preservation reward i.e. ($\gamma=1.0$) has the best trade-off showing significantly improved MP score (p$<$0.0003) without significant drop in WER (p=$0.54$) compared to RLHF-\texttt{\textbf{WER}}  $(\gamma=0.00)$.
Table~\ref{tab:examples_rlhf_wer_mp} shows examples of a few ground truth and predicted transcripts from the two models. We can see from the examples that even on transcripts where both models have high WERs, the signal from meaning preservation ($\gamma=1.0$) improves the final prediction.

\subsection{Ablation: RLHF weights and impact on MP score}
We also show ablations at different values of $\gamma$ in Table~\ref{tab:gemma_rlhf}. Fig.~\ref{fig:rlhf_mp_severity}(b) compares MP score on Euphonia test and dev, and we see that increasing the strength of $\gamma$ when combining MP score in the reward signal, there are substantial improvements in meaning preservation with little decline (insignificant $p=0.54$) in WER.

\subsection{Break down by severity}
\begin{figure}[!htbp]
    \centering
    \begin{subfigure}[t]{0.23\textwidth}
        \centering
        \includegraphics[width=\textwidth]{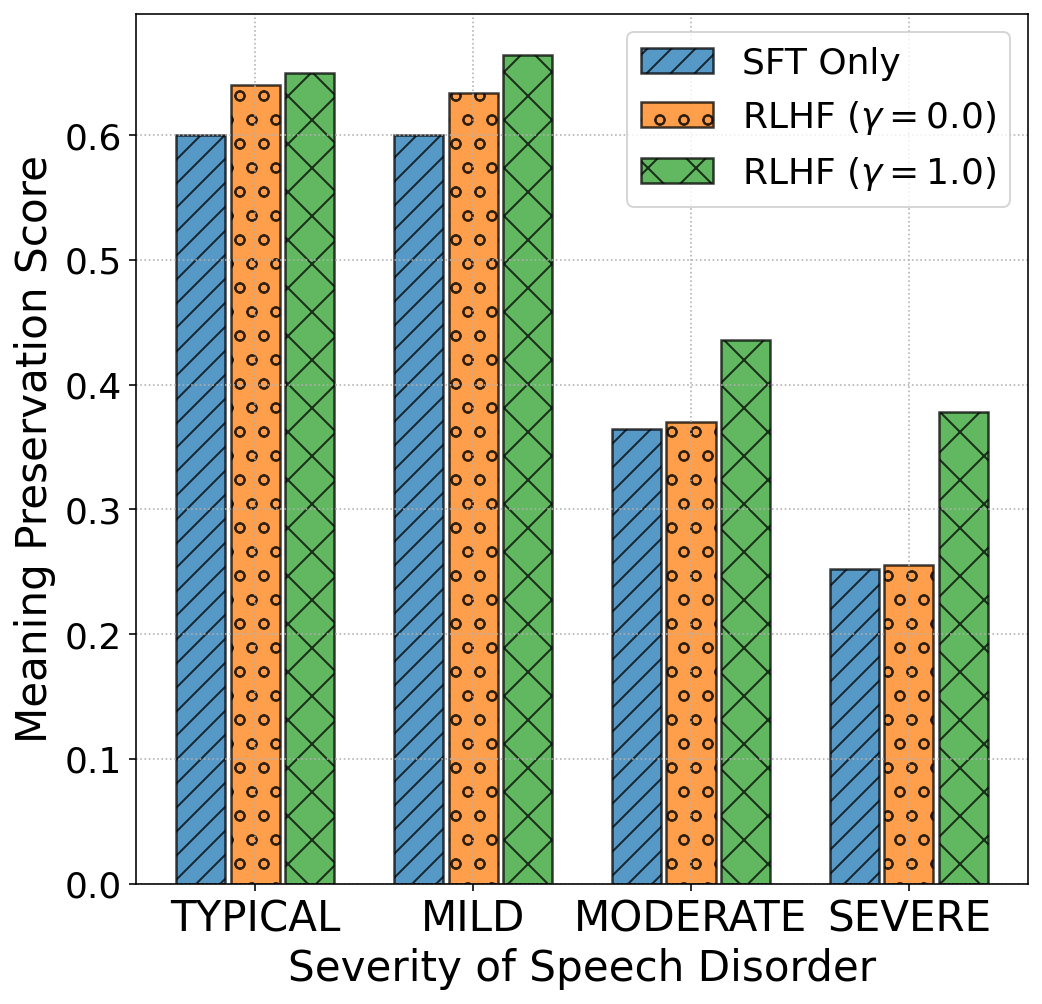}
        \caption{MP score sliced by severity.}
    \end{subfigure}%
    ~ 
    \begin{subfigure}[t]{0.24\textwidth}
        \centering
        \includegraphics[width=\textwidth]{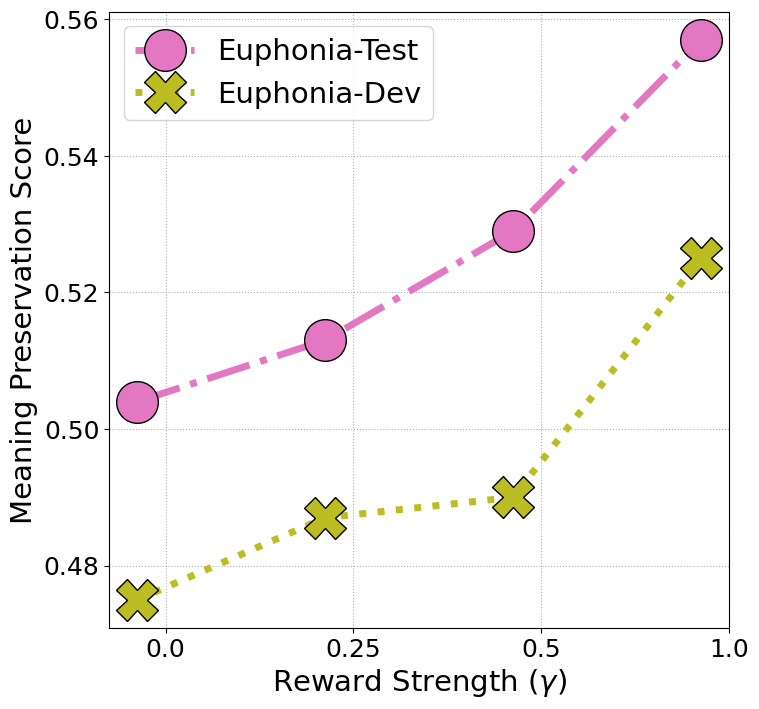}
        \caption{MP score vs strength of $\gamma$.}
    \end{subfigure}
    \caption{Higher strength of $\gamma$ leads to improvement in meaning preservation on the Euphonia val and test sets, and for all severity levels.}
    \label{fig:rlhf_mp_severity}
\end{figure}

Euphonia test set also includes labels for speaker-level severity annotated by Speech and Language Pathologists (SLP). Fig.~\ref{fig:rlhf_mp_severity}(a) compares the model performance on meaning preservation for each severity level and we observe that RLHF models show improved scores across all severity levels, even more so when $\gamma=1$. We also observe larger improvements on individuals with severity ratings as moderate and severe.

\subsection{Human evaluation}

We performed human evaluations, where an SLP assessed the error type and severity of the error (meaning preservation) on a 3-point\footnotemark{}%
scale on 220 examples from the RLHF-\texttt{\textbf{WER}} + \texttt{\textbf{MP}} ($\gamma=0$ and $1.0$) models. \footnotetext{Meaning Completely Preserved, Mild Error, Severe Error}
In Table \ref{tab:human_eval} we list the aggregated statistics considering all errors judged by SLP as \textit{severe} as negative ($-$ve) and the rest as positive ($+$ve). We find that the number of severe errors as judged by the SLP dropped significantly for $\gamma=1.0$ model and that the model based evaluations were significantly correlated ($\rho$=0.684 and $\rho$=0.639) and predictive of the human assessments.

The SLP ratings also supported the analysis in Sec.~\ref{sec:res_rlhf} that the model with $\gamma=1.0$ preserved meaning on more of the utterances. The SLP also identified the types of errors made by both the models and found more normalization errors on the model tuned with MP score weight $\gamma=1.0$. Some examples of these are highlighted in Table~\ref{tab:examples_rlhf_wer_mp}.  %

\begin{table}[!t]
    \centering
    \caption{Human evaluation of model outputs support the analysis from automated metrics.}
    \vspace{-0.6em}
    \begin{tabular}{r|c|c}
    \toprule \midrule
         \textbf{Statistic} (\# samples = 220)& $\gamma=0.0$ & $\gamma=1.0$ \\ \midrule
         \textbf{Average Primary Assessment} (Human MP) & $29.10\%$ & $40.45\%$  \\
         \textbf{Accuracy} (Human vs. Model MP) & $85.90\%$ & $81.36\%$  \\
         \textbf{Spearman} $(\rho)$ (Human vs. Model MP)  & 0.684$^*$ & 0.639$^*$  \\

         \bottomrule
    \end{tabular}
    \label{tab:human_eval}
\end{table}

\section{Summary/Conclusion}

In this paper we proposed a new strategy for ASR adapted to a different domain such as disordered speech based on an existing LLM backbone. Our proposed technique uses a frozen audio encoder and existing LLM architecture, and demonstrates how audio tokens can be mapped to text tokens in the language model and tuned to enable the LLM to recognize speech. We demonstrated that supervised finetuning by adding disordered speech data with a large open-domain parallel speech to text corpus followed by reinforcement learning on just the disordered speech data from the domain led to significant improvements in the model performance relative to standard finetuning on domain specific data. In particular, we demonstrated how reward signals can be obtained on measures of both syntactic and semantic accuracy using a second LLM to predict meaning preservation scores on the generated transcripts. Our results demonstrate a promising approach to adapting LLMs to recognize speech on a different domain using RL. Our experiments were on the open source Gemma model, and it remains as future work to apply this with other similar and larger models, as well as models languages other than English~\cite{jiang2024mixtral,nakamura2024aurora}. %
Developing better audio token discretization strategies and combining multiple reward signals are also interesting future directions.

\newpage

\bibliographystyle{IEEEtran}
\bibliography{refs}

\end{document}